\documentclass[pdflatex, 5p, preprint]{elsarticle}

\usepackage{amsmath} 
\usepackage{amstext}
\usepackage{amssymb}
\usepackage{url}

\def\eqref#1{(\ref{#1})}

\def\o{{\cal O}}

\begin{document}

\begin{frontmatter}
\title{Computation of electron quantum transport in graphene nanoribbons using GPU}
\author{S. Ihnatsenka}
\address{Department of Physics, Simon Fraser University, Burnaby, British Columbia, Canada V5A 1S6}
\tnotetext[t1]{Comput. Phys. Commun. 183 (2012) 543-546}

\begin{abstract}
The performance potential for simulating quantum electron transport on graphical processing units (GPUs) is studied. Using graphene ribbons of realistic sizes as an example it is shown that GPUs provide significant speed-ups in comparison to central processing units as the transverse dimension of the ribbon grows. The recursive Green's function algorithm is employed and implementation details on GPUs are discussed. Calculated conductances were found to accumulate significant numerical error due to single-precision floating-point arithmetic at energies close to the charge neutrality point of the graphene.
\end{abstract}


\end{frontmatter}

During several decades the demands and importance of high-performance computation in physics have been steadily growing. Electron quantum transport simulations are one of the relevant fields. They are usually performed either for small device geometries or greatly reduced basis set of underlying atomic orbitals. Rough approximations, e.g., parabolic dispersion relations or plane wave bases, are frequently employed to compute the electrical conductances and currents.\cite{Datta_book} However, only ab-initio computations are able to recover physical phenomena correctly and assess the validity of simpler approaches. Before the simulation even starts it is nessesarily to compromise between accuracy and computation speed. Such a choice is mainly determined by the available computer resources. At the same time experimental data is usually available for devices whose geometries compose thousands or even millions of atoms. A prominent example of such devices is graphene nanoribbons studied experimentally by Lin etal.\cite{Lin08} Their devices were 30 nm wide and 900 - 1700 nm long. The devices were fabricated from a graphene monolayer that represents a honeycomb lattice of carbon atoms separated by a 0.142 nm distance. This immediately gives 1 - 2 million carbon atoms that need to be taken into account. To treat such large quantities advanced numerical methods should be employed, for example, the recursive Green's function formalism.\cite{Datta_book, disorder} Even though such methods are used one still faces a need for huge computing power. The study presented here explores the possibility of using a Graphical Processing Unit (GPU) for effective computing of quantum electron transport through graphene nanoribbons. 

Over the last few years it has been realized that the large computational power of GPU could be used for purposes other than the video game industry. This power results from the relative simplicity of the GPUs architecture. Its performance exceeds that of Central Processing Unit (CPU) by large factors because of the large number of parallel processing units on a single chip. By design, GPUs are optimized for manipulating a large number of graphics primitives in parallel, which often amounts to simple, floating-point matrix calculations. In contrast to current CPUs, they are not designed to cope with ``unexpected'' branches in the code, or for executing a single-threaded program as fast as possible. While this makes GPUs not well suited as drop-in replacements for CPUs, their highly parallel architecture might be advantageous for scientific calculations with a large part of parallelizable code. Their original design for graphics calculations however entails certain design features which are not necessarily optimal for scientific computational tasks, such as a special hierarchy of memory organization or a restriction to efficient floating-point calculations only in single precision arithmetic. GPU usage for scientific applications became much easier with advent of language extension NVIDIA CUDA.\cite{cuda} The results presented below were obtained using a code written in the C language within the CUDA framework.  

\begin{figure}[tb]
\includegraphics[keepaspectratio,width=\columnwidth]{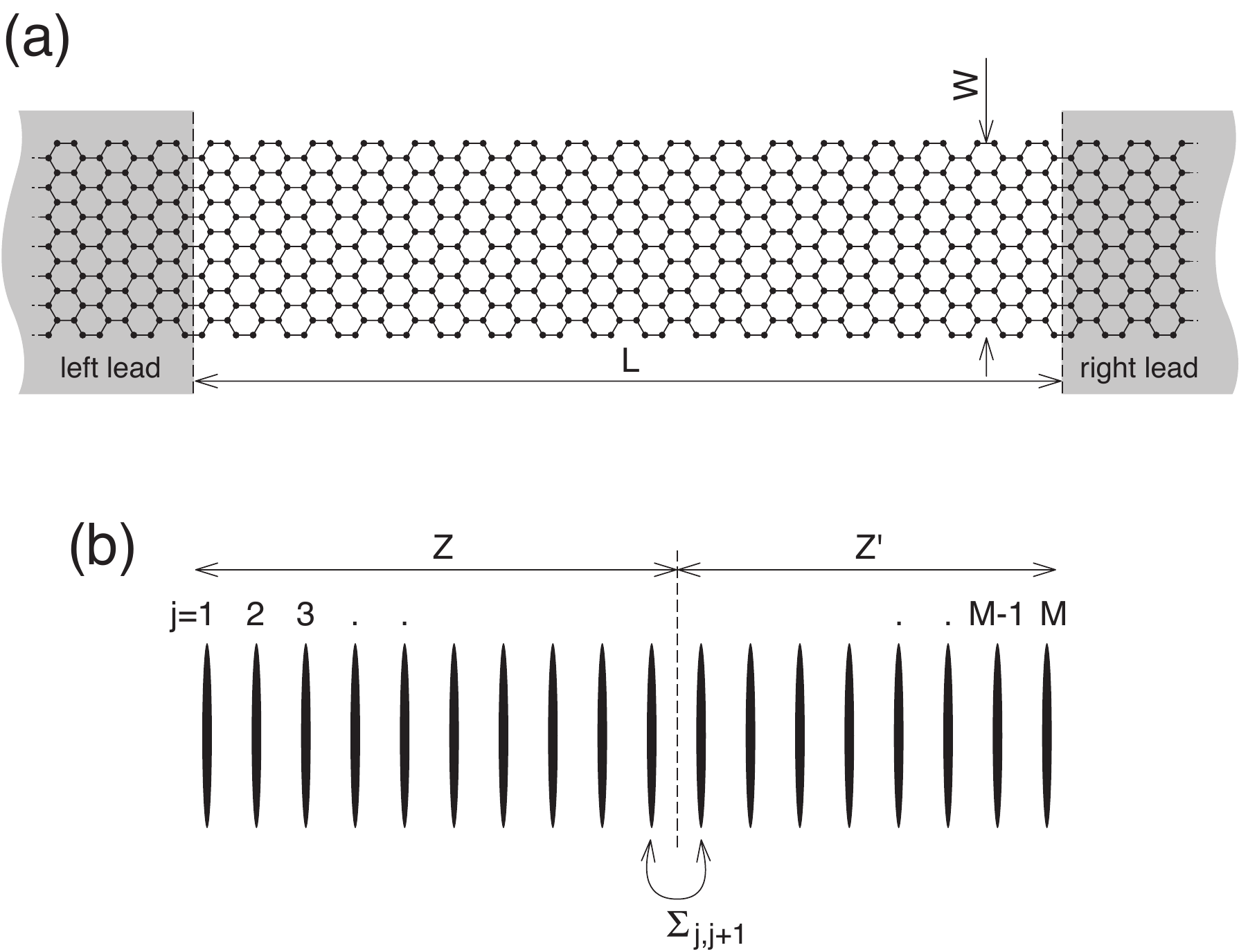}
\caption{(a) Graphene nanoribbon. Honeycomb lattice of carbon atoms forms a strip of width $W$ and length $L$. Semi-infinite leads are attached on both ends of the ribbon. (b) Schematic of the device for application of Dyson's equation by splitting the device in two parts.}
\label{fig:1}
\end{figure}

The structures used for computer simulations are graphene nanoribbons. They are made from a single layer graphene tailored into narrow strips of width $W$, see Fig. 1(a). Numerical computations are explicitly performed for a central scattering region of length $L$. It is attached at its ends to two semi-infinite graphene regions of the same width. These semi-infinite regions serve as electron reservoirs supplying conducting electrons to the system. To simplify the computation each carbon atom is represented by a single $p$ orbital. To a first approximation, this orbital determines the electronic properties of the graphene. The atomic configuration along the edges of the ribbon is armchair for all of the following results.

One of the most effective methods for quantum electron transport computation is the recursive Green's function formalism.\cite{Datta_book, opendot, Zozoulenko96, Anantram08} Its central quantity is the Green's function\cite{Datta_book}
\begin{equation}
  \mathcal{G} = (E - H)^{-1}
  \label{eq:1}
\end{equation}
where $E$ is the electron energy and $H$ is the Hamiltonian of the system. The latter includes the effects of atomic orbitals, leads, different lattice defects and others. Knowledge of the Green's function allows one to compute any desired quantity, e.g. charge density or conductance, that can be directly compared with experimentally measured quantities.\cite{Datta_book} The conductance reads
\begin{equation}
  G = \frac{2e^2}{h} \text{Tr}[\Gamma_L\mathcal{G}^a\Gamma_R\mathcal{G}^r].
  \label{eq:2}
\end{equation}
Here Tr represents the trace and the elements of the matrix $\Gamma_{R(L)}$ couple the leads with the scattering region, see Ref. \cite{Datta_book} for details. The retarded and advanced Green's function $\mathcal{G}^r$ and $\mathcal{G}^a$ are computed by the recursive algorithm\cite{Zozoulenko96, Anantram08} via successive application of the Dyson equation
\begin{equation}
  \mathcal{G} = \mathcal{G}^0 + \mathcal{G}^0\Sigma\mathcal{G}.
  \label{eq:3}
\end{equation}
This equation relates the Green's function of the full system $Z + Z'$ in terms of the subsystems $Z$, $Z'$ and the coupling $\Sigma$ between $Z$ and $Z'$, see Figure 1(b). The algorithm starts by partitioning the scattering region into slices each accounting for the on-site interactions as well as interactions between elements within the given slice. Each slice is represented by a square matrix of size equal to the number of orbitals $N$ (or carbon atoms in the present study). The Green's functions at slice $j$, after simple algebra, can be written as\cite{Anantram08}
\begin{eqnarray}
\mathcal{G}_{jj} &=& (1-\mathcal{G}^0_{jj}\Sigma^l\mathcal{G}_{j-1,j-1}\Sigma^r)^{-1}\mathcal{G}^0_{jj}, \\ 
\mathcal{G}_{1j} &=& \mathcal{G}_{1,j-1}\Sigma^r\mathcal{G}_{jj}.
  \label{eq:4}
\end{eqnarray}
These equations are systematically applied from the first till the last, $M$-th, slice of the scattering region; $\mathcal{G}_{1M} \sim \mathcal{G}^r$. Each recursion loop involves at least 5 matrix multiplications along with one call of the linear equation solver. Computation time depends on both matrix dimension as well as the number of recursion loops, or, in other words, on the dimensions of the scattering region. Note that, at the 1-st and $M$-th slices, one should also find and include the self-energies due to the leads which requires solution of an eigen-problem. The most time consuming part of the algorithm is recursion over the scattering region. One should therefore primarily target optimization of the recursion loops, Eqs. (4)-(5).

To attain high performance of the recursive Green's function algorithm and use the whole power of the GPU one should perform recursive calculations on the GPU alone, without any data transfer to the global (host) computer memory and CPU calls. Matrix-matrix operations involved in each recursion loop can be effectively computed using the highly optimized CUBLAS\cite{cublas} and CULA\cite{cula} libraries. They provide duplicates of the well-known BLAS and LAPACK routines ported to the GPU. In total, a simulation proceeds in the following way:
\begin{itemize}       
\item Input data, define coupling self-energies, generate atomic defects and other scattering parameters if any.
\item Compute self-energies due to the leads.
\item Allocate GPU memory, transfer data from the host memory to the device memory.
\item Recursion loops: Build up the Hamiltonian for each slice and compute $\mathcal{G}_{jj}$, $\mathcal{G}_{1j}$, Eqs. (4)-(5).
\item Transfer the computed Green's functions for the whole scattering region back to the host memory. Deallocate GPU memory.
\item Calculate the conductance, Eq. \ref{eq:2}, output the results.
\end{itemize}


\begin{figure}[tb]
\includegraphics[scale=0.7]{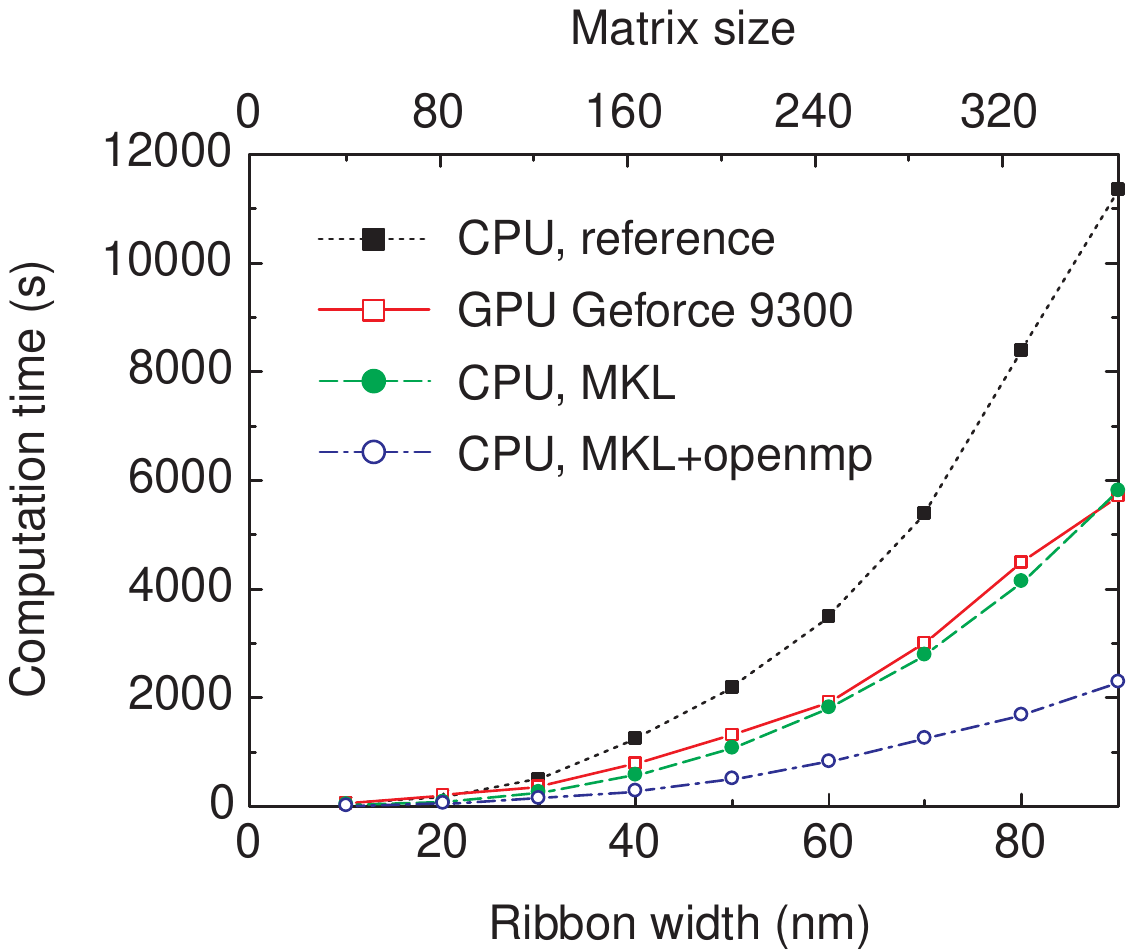}
\caption{Computation times for the conductance of the graphene ribbon vs. the ribbon width or matrix size. The matrix size is equal to number of carbon atoms in the cross section. The length of the ribbon is fixed $L=1700$ nm.  The dotted line with filled squares shows CPU timings with reference (unoptimized) BLAS and LAPACK libraries. The solid line with open squares corresponds to GPU GeForce 9300. The dash-dotted lines with filled and open circles show CPU timings with Intel MKL for single and multi-threaded modes, respectively. The multi-threaded mode is implemented within Open MP framework. CPU is 3.0 GHz Intel Core Quad.}
\label{fig:2}
\end{figure}

Figure \ref{fig:2} shows computation times for the conductance in the graphene ribbons as a function of the ribbon width. The ribbon length is kept constant as $L=1700$ nm that equals the device size experimentally studied in Ref. \cite{Lin08}. Note that the time needed for computation of the lead self-energies, the second step in above step flow, is excluded from the data in Fig. \ref{fig:2} because it is always performed on the CPU and takes less than 0.3$\%$ of total computation time. The solid line with open squares shows timings for GPU GeForce 9300, which is a low-cost hardware solution. The dotted line with filled squares corresponds to a single core of a 3.0 GHz Intel Core Quad along with unoptimized reference BLAS and LAPACK implementation.\cite{Blackford, Anderson} In general, the CPU performs faster for narrow ribbons, $W<10$ nm, but becomes inferior to the GPU as the ribbon widens. This trend agrees with other results reported in literature.\cite{GPU_other} For narrow ribbons, data transter turns out to be a bottleneck though massive matrix operations become a limiting factor for wider devices. 

The GPU GeForce 9300 used in the present simulations has 2 multiprocessors each boarding 8 cores. The latter operate at 1.2 GHz. Because the GPU is used also for graphic needs the operating system allocates only a fraction of the resources for computations. As a result the GPU performance doesn't approach maximum values: it achieves roughly half of the theoretical processing rate. Nevertheless it is still twice as fast as a single core of the 3.0 GHz Intel Core Quad for wide graphene ribbons, Figure \ref{fig:2}. Using a dedicated general purpose GPU, like the recent Tesla C1060 card, will substantially improve the computation rate. 

The CPU performance might be improved by using multi-threaded libraries like the Intel Math Kernel Library (MKL). The latter includes BLAS and LAPACK routines optimized for Intel CPUs. The results for both single and multi-threaded MKL are shown in Figure \ref{fig:2}. Computation time drops by several times in comparison to reference implementation of BLAS and LAPACK on Netlib. This gain is obtained by minor efforts. Another improvement might be achieved via MPICH parallelization that should theoretically speed-up by a factor of $\sim$4 for the Intel Core Quad when compared to single-core (or single-treaded) code implementation. This however will require modification of the recursion algorithm, Eqs. (4)-(5). Substantial modifications of the recursion algorithm will also be needed to automate (non optimized) parallel core calculations by a compiler itself.

\begin{figure}[tb]
\includegraphics[scale=0.7]{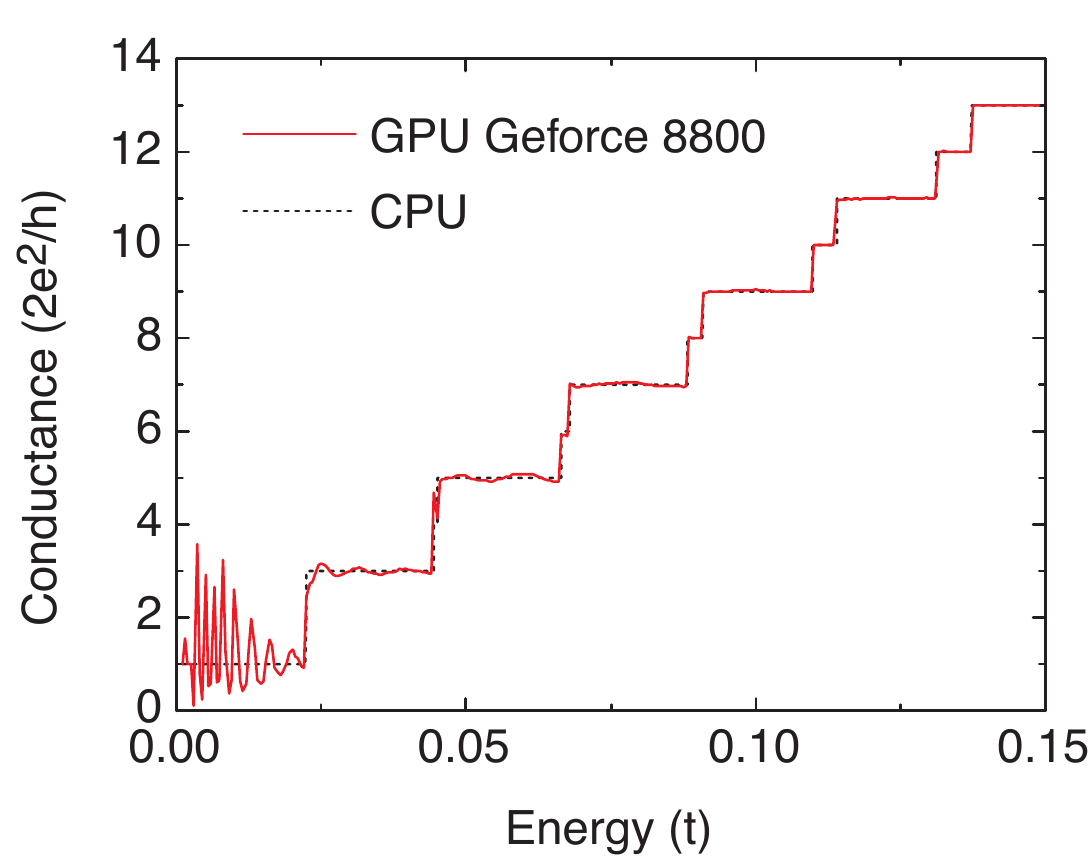}
\caption{The conductance as a function of the Fermi energy for the graphene nanoribbon of $W=30$ nm and $L=1700$ nm. The conductances calculated on the GPU deviate from the exact values obtained on the CPU that uses double-precision arithmetic. The GPU computation uses single-precision floating-point arithmetic.}
\label{fig:3}
\end{figure}

An important issue of GPU computation is its accuracy since single-precision floating-point arithmetic is used to achieve the greatest efficiency. This leads to numerical errors systematically accumulated during recursion that exceed the corresponding numerical error on the CPU where double-precision arithmetic is used. The latter actually might be considered as an exact result because CPU accuracy was found to be better than $10^{-9}$. Note that the CPU performance doesn't change with the precision of the arithmetic. Figure \ref{fig:3} compares the conductances for an ideal graphene ribbon of $W=30$ nm and $L=1700$ nm for the GPU and CPU architectures. In ideal case, they should reveal a step-like dependence with the conductance being a multiple of $\frac{2e^2}{h}$. However, the GPU conductances show strong irregular oscillations at low energies. These deviations are caused by (i) the numerical error incurred for the quantum-mechanical system with many evanescent states and (ii) particular structure of the unconfined wave function at the first electronic subband.\cite{Yamamoto09} The numerical error might exceed 200$\%$ at low energies though it quickly decreases when conduction is dominated by three or more propagating states, i.e. $G>3\times \frac{2e^2}{h}$. It turns out that the main source of the error is the linear equation solver (CGESV) in the current implementation of the recursive Green's function algorithm. As soon as GPUs will operate with double-presicion arithmetic, which has became supported in a recent generation of GPU cards, the numerical error is expected to descrease and the computation results to match the precise results. It worth finally noting that double-precision computation performs slower than single-precision one.

In conclusion, the present study showed that electron quantum transport simulations might be effectively done using the GPU architecture that outperforms the CPU even for a low-cost build-in hardware solution. Computational speed-ups might be higher by large factors for ``cost-comparable'' hardware. The recursive Green's function algorithm can be straightforwardly implemented in the CUDA framework. Highly optimized CUBLAS and CULA libraries allow one-to-one replacement of BLAS and LAPACK libraries. However, because the GPU architecture operates on single-precision floating-point data substantial numerical error might accumulate.

The author is grateful to G. Kirczenow for critical reading of the manuscript.


\end{document}